\documentstyle[newlfont,amssymb,prd,
preprint,
tighten,aps]{revtex}
\begin{document}
\draft
\widetext
\preprint{\mbox{
\begin{tabular}{l}
KIAS-P97003\\
DFTT 55/97\\
hep-ph/9709494\\
September 1997
\end{tabular}
}}
\title{When do neutrinos cease to oscillate?}
\author{C. Giunti$^{\mathrm a}$,
        C. W. Kim$^{\mathrm b,c}$,
        and U. W. Lee$^{\mathrm d}$}
\address{$^{\mathrm a}$
INFN, Sez. di Torino, and Dip. di Fisica Teorica,
Univ. di Torino, I--10125 Torino, Italy\\
         $^{\mathrm b}$
         Department of Physics $\&$ Astronomy,
         The Johns Hopkins University,
         Baltimore, MD 21218\\
         $^{\mathrm c}$
         School of Physics, Korea Institute for Advance Study,
         Seoul 130-012, Republic of Korea \\
         $^{\mathrm d}$
         Department of Physics,
         Mokpo National University,
         Chonnam 534-729,
         Republic of Korea
         }
\date{September 29, 1997}
\maketitle
\begin{abstract}
In order to investigate when neutrinos cease to oscillate in the
framework of quantum field theory,
we have reexamined the wave packet treatment
of neutrino oscillations
by
taking different sizes of the wave packets
of the particles involved
in the production and detection processes.
The treatment is shown to be considerably simplified
by using the Grimus--Stockinger theorem
which enables us to carry out the integration
over the momentum of the propagating neutrino.
Our new results
confirm the recent observation
by Kiers, Nussinov and Weiss
that
a precise measurement of the energies
of the particles involved in the
detection process would increase
the coherence length.
We also present a precise definition of the coherence length
beyond which neutrinos cease to oscillate.
\end{abstract}


%
%


A rigorous treatment of neutrino oscillations requires
the study of the processes in which
neutrinos are produced and detected
\cite{Nussinov,BP78,Kayser,GKL91,GKLL93,Rich,KNW96,GS96,Campagne}.
Following the first attempts
\cite{Nussinov,BP78,Kayser}
to develop
a proper use of the wave packet formalism,
we have carried out
detailed calculations
in both quantum mechanics \cite{GKL91}
and quantum field theory \cite{GKLL93},
with a quantitative derivation of the coherence length
for neutrino oscillations.
Due to the complexity of calculations,
however,
in \cite{GKLL93} it was assumed
that the sizes of the wave packets of
the initial and final particles are all the same.

Recently,
Grimus and Stockinger \cite{GS96} proved
an elegant and very useful theorem
that allows to simplify the wave packet treatment of
neutrino oscillations.
Taking advantage of this theorem,
we have re-derived the neutrino oscillation formula
in the general case in which
all the particles involved in the production and detection
processes have different wave packet sizes.
Our new result confirms in the framework of a
quantum field theoretical approach
an interesting observation presented in \cite{KNW96}
that
an accurate measurement
of the energies of the particles involved
in the detection process leads to an increase of
the coherence length for neutrino oscillations.


Let us consider the flavor-changing process
\begin{eqnarray}
P_I \to P_F + \ell^{+}_{\alpha} + \null & \nu_{\alpha} & \nonumber \\
& \downarrow & \scriptstyle ( \nu_{\alpha} \to \nu_{\beta} ) \nonumber \\
&\nu_{\beta} & \null + D_I \to D_F + \ell^{-}_{\beta} \,,
\label{01}
\end{eqnarray}
where $ P_{I} $ and $ P_{F} $ ($ D_{I} $ and $D_{F} $)
are the initial and final production (detection) particles.
The process (\ref{01})
takes place through the intermediate propagation of a neutrino,
which oscillates  from flavor $\alpha$
to flavor $\beta$  (here $\alpha,\beta=e, \mu, \tau$).
In the process (\ref{01}),
the production and detection interactions
are  localized at the coordinates
$ (\vec{X}_{P} , T_{P})$
and
$ ( \vec{X}_{D} , T_{D} ) $, respectively.

The form of the wave functions of the initial and final particles
involved in the process (\ref{01}) is determined by
how the initial particles are prepared
and how the final particles are detected.
In the following, we will assume, for simplicity,
Gaussian wave functions, whose wave packet forms
in momentum and coordinate space are given in \cite{GKLL93}.
The wave packets in momentum space are assumed to be sharply peaked
around their average momenta, which are denoted by
$
\langle\vec{p}_{k}\rangle
$,
where
$
k=P_I,P_F,\alpha,D_I,D_F,\beta
$.
The corresponding average energies
$
\langle{E_{k}}\rangle
$
are given by
$
\langle{E_{k}}\rangle
=
\sqrt{ \langle\vec{p}_{k}\rangle^2 + m_{k}^2 }
$,
where $m_{k}$ is the mass of the $k^{\mathrm{th}}$ particle.
In order to make a realistic calculation,
we will consider a different
spatial width $\sigma_{x{k}}$
for the wave packet of each particle
involved in the process (\ref{01}).
Let us emphasize that
the localization of the particles
does not require necessarily
the action of a man-made apparatus,
but can be determined by the environment
in which the process (\ref{01}) takes place.
 
Following the method presented in \cite{GKLL93},
the amplitude of the process (\ref{01}) 
can be written as (see Eq.(8) of \cite{GKLL93})
%
%
\begin{eqnarray}
&&
A_{\alpha\beta}(\vec{L},T)
\propto
\sum_{a}
\mathcal{U}_{{\alpha}a}^{*} \,
\mathcal{U}_{{\beta}a} 
\int
\frac{ \mathrm{d}^4 q }{ (2\pi)^4 } \,
\overline{U}_{D} \,
\frac{ {q\kern-.46em/} }
     { q^2 - m_{a}^2 + i \epsilon } \,
V_{P} \,
\exp\left[ - i q_0 T + i \vec{q} \cdot \vec{L} \right]
\nonumber \\
&&
\null \hspace{0.5cm}
\times
\int \mathrm{d}^4 x_1 \, 
\exp\left[ - i \left( E_P - q_0  \right) t_1 
+ i \left( \vec{p}_P - \vec{q} \right) \cdot \vec{x}_1
-
\frac{ \vec{x}_1^2 - 2 \, \vec{v}_{P} \cdot \vec{x}_1 \, t_1
+ \Sigma_{P} \, t_1^2 }
{ 4 \sigma_{xP}^2 }
\right] \nonumber \\
&&
\null \hspace{0.5cm}
\times
\int \mathrm{d}^4 x_2 \,
\exp \left[  - i \left( E_D + q_0 \right) t_2 
+ i \left( \vec{p}_D + \vec{q} \right) \cdot \vec{x}_2
-
\frac{ \vec{x}_2^2 - 2 \, \vec{v}_{D} \cdot \vec{x}_2 \, t_2
+ \Sigma_{D} \, t_2^2}
{ 4 \sigma_{xD}^2 }
\right] \,,
\label{02}
\end{eqnarray}
%
%
where
the index $a$ denotes the neutrino field
$\nu_a$ with mass $m_a$,
$\mathcal{U}$ is the neutrino mixing matrix,
$
\vec{L} \equiv \vec{X}_D - \vec{X}_P
$
is the macroscopic distance vector
from the neutrino production point
to the neutrino detection point
and
$
T \equiv T_D - T_P
$
is the macroscopic time interval $T$
between neutrino production and detection.
In Eq.(\ref{02})
we have defined various quantities
relative to the production process
as follows
(the corresponding quantities relative to the detection process
are denoted with $P$ replaced by $D$
and $\alpha$ replaced by $\beta$
in the subscript).
$ E_{P} $ and $ \vec{p}_{P} $
are given by
\begin{eqnarray}
&& E_{P} \equiv \langle E_{P_I} \rangle -
   \langle E_{P_F}  \rangle -
   \langle E_{\alpha} \rangle \,, \\
&& \vec{p}_{P} \equiv
          \langle \vec{p}_{P_I} \rangle
        - \langle \vec{p}_{P_F} \rangle
        - \langle \vec{p}_{\alpha} \rangle \,.
\end{eqnarray}
The quantity $\sigma_{xP}$,
given by
\begin{equation}
\frac{1}{\sigma_{xP}^2} \equiv
     \frac{1}{\sigma_{xP_{I}}^2}
   + \frac{1}{\sigma_{xP_{F}}^2}
   + \frac{1}{\sigma_{x\alpha}^2}
   \,,
   \label{sigmaxP} 
\end{equation}
can be considered as the effective size of the
production process.
Notice that
the width of the wave packet of
the most localized particle
(\emph{i.e.} the one with the smallest $\sigma_x$)
dominates in the determination of $ \sigma_{xP} $.
For example, in \cite{KNW96}
it is argued that the width $\sigma_{xe}$ of the captured electron
in the production process of $^7\mathrm{Be}$ solar neutrinos,
$
e^{-} + \null^7\mathrm{Be} \to \null^7\mathrm{Li} + \nu_{e}
\,,
$
is much smaller than the widths of the wave packets
of the other particles involved in the process.
Therefore, in this case
$\sigma_{xP}\simeq\sigma_{xe}$
(the formalism presented here
can be obviously generalized
for the case of more than one initial particle).

In a quantum-mechanical framework
(see \cite{Nussinov,BP78,Kayser,GKL91,KNW96})
the size of the production process determines
the size of the wave packet of the propagating neutrino.
On the other hand,
in the approach adopted here
the neutrino is treated as a
``virtual''
particle propagating
between the production and detection vertices,
whose properties are defined by
the production and detection interactions
in an equal way.
Hence,
any attempt to define
a neutrino wave packet
in the framework of the calculation presented here
would lead to a definition of its size
in equal terms of both
$\sigma_{xP}$ and $\sigma_{xD}$.
Such an interpretation is unacceptable
from a causal point of view.
However,
the present approach
does not need an interpretation in terms
of neutrino properties,
because the neutrino is not directly observed,
and
we will show that,
nevertheless,
it leads to a rigorous derivation
of the oscillation probability,
including the effect of the coherence length.

The quantities
$ \vec{v}_{P} $ and $ \Sigma_{P} $
in Eq.(\ref{02})
are given by
\begin{eqnarray}
&&
\vec{v}_{P} \equiv \sigma_{xP}^2
\left(
  \frac{ \vec{v}_{P_I} }{ \sigma_{xP_{I}}^2 }
+ \frac{ \vec{v}_{P_F} }{ \sigma_{xP_{F}}^2 }
+ \frac{ \vec{v}_{\alpha} }{ \sigma_{x\alpha}^2 }
\right) ,
\label{vP}
\\
&&
\Sigma_{P} \equiv \sigma_{xP}^2
\left(
  \frac{ \vec{v}_{P_I}^2 }{ \sigma_{xP_{I}}^2 }
+ \frac{ \vec{v}_{P_F}^2 }{ \sigma_{xP_{F}}^2 }
+ \frac{ \vec{v}_{\alpha}^2 }{ \sigma_{x\alpha}^2 }
\right) ,
\label{SP}
\end{eqnarray}
where
$
\vec{v}_{k} \equiv
        {\langle\vec{p}_{k}\rangle} \,
        / \, {\langle{E_{k}}\rangle} \,
$
is the group velocity of the $k^{\mathrm{th}}$ particle.
Finally, the definitions of
$
\overline{U}_{D}$ and $V_{P}
$
are given by
\begin{eqnarray}
&&
\overline{U}_{D} \equiv
J^{D}_{\rho}(\langle\vec{p}_{D_F}\rangle,\langle\vec{p}_{D_I}\rangle) \,
\overline{u}_{\beta}(\langle\vec{p}_{\beta}\rangle) \,
\gamma^{\rho} (1+\gamma_{5}) \,,
\label{740272} \\
&&
V_{P} \equiv (1+\gamma_{5}) \gamma^{\rho} \,
v_{\alpha}(\langle\vec{p}_{\alpha}\rangle) \,
J^{P}_{\rho}(\langle\vec{p}_{P_F}\rangle,\langle\vec{p}_{P_I}\rangle) \,,
\label{740271} 
\end{eqnarray}
where
$
J^{P}_{\rho}(\langle\vec{p}_{P_F}\rangle,\langle\vec{p}_{P_I}\rangle)
$
and
$
J^{D}_{\rho}(\langle\vec{p}_{D_F}\rangle,\langle\vec{p}_{D_I}\rangle)
$
are the matrix elements of the weak currents of
the production and detection particles. 

%
%
Carrying out the gaussian integrals over $x_1$ and $x_2$, we obtain
\begin{eqnarray}
A_{\alpha\beta}(\vec{L},T)
&\propto&
\sum_{a} \mathcal{U}_{{\alpha}a}^{*} \, \mathcal{U}_{{\beta}a}
\int \frac{ \mathrm{d}^4 q }{ (2\pi)^4 } \, \overline{U}_{D} \,
\frac{ {q\kern-.46em/} }
     { q^2 - m_{a}^2 + i \epsilon } \,  V_{P} \,
\exp\left[ - i q_0 T + i \vec{q} \cdot \vec{L} \right]
\nonumber \\
&&
\hskip2cm
\null \times
\exp \left[ - \frac{ \left( \vec{p}_P - \vec{q} \right)^2 }
       { 4 \, \sigma_{pP}^2 }
- \frac{ \left[ \left( E_P - q_0  \right) -
             \left( \vec{p}_P - \vec{q} \right)
             \cdot \vec{v}_{P} \right]^2 }
       { 4 \, \sigma_{pP}^2 \lambda_P }
            \right.
\nonumber \\
&&
\hskip2cm
\phantom{ \times \exp[[ }
\left. \null
             - \frac{ \left( \vec{p}_D + \vec{q} \right)^2 }
                    { 4 \, \sigma_{pD}^2 }
             - \frac{ \left[ \left( E_D + q_0 \right) -
                          \left( \vec{p}_D + \vec{q} \right)
                          \cdot \vec{v}_{D} \right]^2 }
                    { 4 \, \sigma_{pD}^2 \lambda_D }
\right] \,,
\label{7404}
\end{eqnarray}
where we have defined
$
\lambda_{P(D)} \equiv \Sigma_{P(D)} - \vec{v}_{P(D)}^2
$
and
$\sigma_{pP(D)}$
is defined by the relation
$
\sigma_{xP(D)} \, \sigma_{pP(D)} = 1/2
$.
Notice that the integrations over
$x_1$
and
$x_2$
in Eq.(\ref{02})
did not produce the usual
$\delta$-functions representing
energy--momentum conservation at each interaction vertex.
Indeed,
in the wave packet treatment the
energies and momenta of the particles
involved in the process under consideration
do not have a precise value,
allowing for the uncertainty in energy--momentum conservation
that is necessary for the occurrence of neutrino oscillations
(see \cite{Kayser,GKL91,GKLL93}).

From Eqs.(\ref{vP}) and (\ref{SP})
one can see that
$ 0 \leq \lambda_{P(D)} < 1 $.
Hence,
it is natural to ask what happens if
$ \lambda_P = 0 $
and/or
$ \lambda_D = 0 $.
Let us consider, for simplicity,
the case of only
$ \lambda_P = 0 $.
This situation could arise,
for example,
if only the initial particle $P_I$
in the production process
is observed with a good localization
at $\vec{X}_P$ and
is described by a wave packet,
whereas all the particles in the final state
are described by plane waves.
In this case we have
$\sigma_{xP} =  \sigma_{xP_I}$,
$\vec{v}_{P} =  \vec{v}_{P_I}$
and
$\Sigma_{P}={\vec{v}_{P_I}}^2$,
leading to
$\lambda_D = 0$.
From Eq.(\ref{02})
one can see that
$\Sigma_{P}={\vec{v}_{P_I}}^2$
implies exact energy conservation
in the production process
($\delta(q_0-E_P)$).
If $\vec{v}_{P}\neq0$,
there is also exact momentum conservation
in the production process
($\delta^3(\vec{q}-\vec{p}_P)$)
and neutrino oscillations do not occur
because exact energy--momentum conservation
in the production process can be satisfied only
by one of the neutrino mass eigenstates,
excluding the coherent production
of more than one neutrino mass eigenstates
that is
necessary for neutrino oscillations.
This fact is also clear in coordinate space:
if $\vec{v}_{P}\neq0$
only the trajectory of $P_I$ is known,
which is not sufficient to determine
the position $\vec{X}_P$
and the time $T_P$
of the production process.
Hence,
the distance $\vec{L}$ and the time interval $T$
are not defined
and oscillations are not observable.
On the other hand,
in the special case
$\vec{v}_{P}=0$
there is no exact momentum conservation
in the production process
(although energy is exactly conserved)
and neutrino oscillations are observable.
Physically this situation corresponds
to have the initial particle $P_I$
at rest,
in such a way that
$\vec{X}_P$ is known and
the distance $\vec{L}$
is defined,
although $T_P$ and $T$ are not known 
(similar scenarios are discussed in \cite{Rich,GS96}).
However,
as stated above,
the calculation presented here is based on the assumption
that all the initial and final particles
in the process (\ref{01})
are observed and their wave packets are sharply peaked
around the corresponding average momentum.
Therefore,
the case
$ \lambda_P = 0 $
(and/or
$ \lambda_D = 0 $)
corresponds to a different physical process
from the one under consideration,
which can be discussed modifying the calculation
presented here in an appropriate way.
Under our assumptions,
$ \lambda_P $
and
$ \lambda_D $
are dimensionless quantities of order
$10^{-1}$.
For example,
if the production process is pion decay at rest,
we have
$
\lambda_P
=
\frac{\sigma_{xP}^2}{\sigma_{x\alpha}^2}
\left(
1
-
\frac{\sigma_{xP}^2}{\sigma_{x\alpha}^2}
\right)
\vec{v}_{\alpha}^2
$
with
$
\vec{v}_{\alpha}^2
=
(m_{\pi}^2-m_{\alpha}^2)^2
/
(m_{\pi}^2+m_{\alpha}^2)^2
$,
\emph{i.e.}
$ \vec{v}_{\alpha}^2 \simeq 7 \times 10^{-2} $
for $\alpha=\mu$
and
$ \vec{v}_{\alpha}^2 \simeq 1 $
for $\alpha=e$.

Resuming the calculation of the amplitude
(\ref{7404}),
we perform the integral over $\vec{q}$ using the
Grimus--Stockinger theorem \cite{GS96}:
\begin{equation}
\int \mathrm{d}^3q \,
\frac{ \phi(\vec{q}) \, \mathrm{e}^{ i \vec{q} \cdot \vec{L} } }
     { q_{a}^2 - \vec{q}^2 + i \epsilon }
\ \stackrel{L\to\infty}{\longrightarrow} \
-  \frac{ 2\pi^2 }{ L } \, \phi(q_{a}\vec{\ell}) \,
\mathrm{e}^{ i q_{a} L }
\,,
\label{74501}
\end{equation}
with
$
L \equiv |\vec{L}| $, $ \vec{\ell} \equiv \vec{L} / L
$
and
$
q_{a} = \sqrt{ {q^{0}}^2 - m_{a}^2 }
$.
This theorem is valid 
for a function
$\phi$ which is differentiable
at least three times
such that
$\phi$
itself and its first and second derivatives decrease
at least as
$  1 / \vec{q}^2 $ as $|\vec{q}| \to \infty $.
This is precisely our case.
The remaining integration over
$q^0$
can be done with a saddle-point approximation
at $q^0 = E_a$,
with
$E_a$ given by the relation
\begin{eqnarray}
E_{a}
\,
\frac
{ p_{a} - \vec{\ell} \cdot \vec{p}_P }
{ \sigma_{pP}^2}
-
\frac{
\left( p_{a} - E_{a} \, \vec{\ell} \cdot \vec{v}_P \right)
\left[
\left( E_P - E_{a}  \right)
-
\left( \vec{p}_P - p_{a} \vec{\ell} \right)
\cdot \vec{v}_{P}
\right]
}{ \sigma_{pP}^2 \lambda_P }
&&
\nonumber
\\
\null
+
E_{a}
\,
\frac
{ p_{a} + \vec{\ell} \cdot \vec{p}_D }
{ \sigma_{pD}^2}
+
\frac{
\left( p_{a} - E_{a} \, \vec{\ell} \cdot \vec{v}_D \right)
\left[
\left( E_D + E_{a}  \right)
-
\left( \vec{p}_D + p_{a} \vec{\ell} \right)
\cdot \vec{v}_{D}
\right]
}{ \sigma_{pD}^2 \lambda_D }
&=&
0
\,,
\label{Ea}
\end{eqnarray}
where
$ p_{a} \equiv q_{a}(E_{a}) = \sqrt{E_{a}^2-m_{a}^2} $.
The quantities
$E_{a}$
and
$p_{a}$
can be interpreted as the effective energy and momentum of the
$a^{\mathrm{th}}$
neutrino mass eigenstate
propagating between the production and detection vertices.
In order to simplify our discussion,
in the following we omit explicit terms of the order
$m_a^2 /E^2$,
which are negligible for relativistic neutrinos.
Thus, for $ L \to \infty $
we have
%
%
\begin{equation}
A_{\alpha\beta}(\vec{L},T) \propto
\frac{1}{L}
\sum_{a}
\mathcal{U}_{{\alpha}a}^{*} \,
\mathcal{U}_{{\beta}a} \,
\mathcal{A}_a \,
\frac{1}{\sqrt{\Omega_{a}}} \,
\exp\left[ - i E_a T + i p_a L
- S_{a}\left(E_a\right)
- \frac{1}{2} \,
  \frac { \left( L - v_{a} T \right)^2 }
        { v_{a}^2 \Omega_{a} }
\right]
\label{74508}
\end{equation}
where
$ v_{a} \equiv p_{a} / E_{a} $,
\begin{eqnarray}
\mathcal{A}_a 
&\equiv&
\overline{U}_{D} \left( E_{a} \, \gamma^{0} 
- p_{a} \, \vec{\ell} \cdot \vec{\gamma} 
\right) V_{P} \,,
\label{Aa}
\\
\Omega_{a}
&\equiv&
\frac{1}{ 2 v_{a}^2 }
    \left\{\frac{1}{ \sigma_{pP}^2 }
+ \frac{1}{ \sigma_{pD}^2 }
+ \frac{ \left( v_{a} - \vec{\ell} \cdot \vec{v}_{P} \right)^2 }
       { \sigma_{pP}^2 \lambda_P }
+ \frac{ \left( v_{a} - \vec{\ell} \cdot \vec{v}_{D} \right)^2 }
       { \sigma_{pD}^2 \lambda_D }
\right\}
+ O\left( \frac{m_a^2}{E_a^2} \right)
\,,
\label{Oa}
\\
S_{a}(E_a)
&\equiv&
\frac{ \left( \vec{p}_P - p_{a} \vec{\ell} \right)^2 }
     { 4 \sigma_{pP}^2 }
+
\frac{  \left[ \left( E_P - E_a  \right) -
    \left( \vec{p}_P - p_{a} \vec{\ell} \right)
    \cdot \vec{v}_{P} \right]^2 }
     { 4 \sigma_{pP}^2 \lambda_P }
 \nonumber \\
&& \null
+ \frac{ \left( \vec{p}_D + p_{a} \vec{\ell} \right)^2 }
       { 4 \sigma_{pD}^2 }
+ \frac{ \left[ \left( E_D + E_a \right) -
                \left( \vec{p}_D + p_{a} \vec{\ell} \right)
                \cdot \vec{v}_{D}
          \right]^2 }
       { 4 \sigma_{pD}^2 \lambda_D } \,.
\label{Sa}
\end{eqnarray}

The probability of the process (\ref{01})
is proportional to
$ | A_{\alpha\beta}(\vec{L},T) |^2 $,
but in a practical experimental setting
$\vec{L}$ is usually a fixed and known quantity,
whereas $T$ is not measured.
Therefore, the oscillation probability
at a given distance $\vec{L}$
is given by the time average of
$ | A_{\alpha\beta}(\vec{L},T) |^2 $,
which leads to
\begin{eqnarray}
P_{\alpha\beta}(\vec{L})
&\propto&
\frac{1}{L^2}
\sum_{a,b}
\mathcal{A}_{a} \,
\mathcal{A}_{b}^{*} \,
\mathcal{U}_{{\alpha}a}^{*} \,
\mathcal{U}_{{\beta}a} \,
\mathcal{U}_{{\alpha}b} \,
\mathcal{U}_{{\beta}b}^{*} \,
\frac{1}{\sqrt{\Omega_{a}+\Omega_{b}}}
\,
\exp\left[
- S_{a}\left(E_a\right)
- S_{b}\left(E_b\right)
\right]
\nonumber
\\
&&
\hskip1cm
\null
\times
\exp\left[
- i
\left(
\left( E_a - E_b \right)
\frac
{ v_{a} \Omega_{a} + v_{b} \Omega_{b} }
{ v_{a} v_{b} \left( \Omega_{a} + \Omega_{b} \right) }
- \left( p_a - p_b \right)
\right)
L
\right]
\nonumber
\\
&&
\hskip1cm
\null
\times
\exp\left[
-
\frac{L^2}{2}
\,
\frac
{ \left( v_{a} - v_{b} \right)^2 }
{ v_{a}^2 v_{b}^2 \left( \Omega_{a} + \Omega_{b} \right) }
-
\frac{1}{2}
\left( E_{a} - E_{b} \right)^2
\frac
{ \Omega_{a} \Omega_{b} }
{ \Omega_{a} + \Omega_{b} }
\right]
\,.
\label{exact}
\end{eqnarray}
The overall factor $1/L^2$
represents the geometrical decrease of the neutrino flux.
Notice that the probability
(\ref{exact})
depends not only on the modulus $L$ of the distance,
but also on its direction $\vec{\ell}$,
which is contained in
$\mathcal{A}_{a}$,
$\Omega_{a}$
and
$S_{a}\left(E_a\right)$
(see Eqs.(\ref{Aa})--(\ref{Sa}))
and determines $E_a$ through Eq.(\ref{Ea}).
This is due to the fact that the wave packets of
the external particles are assumed to be known
and the term
$
\exp\left[
- S_{a}\left(E_a\right)
- S_{b}\left(E_b\right)
\right]
$
in Eq.(\ref{exact})
guarantees that the probability of the process
(\ref{01})
is not negligible only when both
$\vec{p}_P$ and $\vec{p}_D$
are aligned with $\vec{\ell}$
within the uncertainty
allowed by the sizes of the wave packets.

Since we are concerned with relativistic neutrinos,
we approximate
\begin{equation}
E_{a} \simeq  E + \rho \, \frac{m_{a}^2}{2E}
\,,
\label{E}
\end{equation}
with $E$ and $\rho$
determined,
respectively,
by the expansion of Eq.(\ref{Ea})
at zeroth and first order in powers of
$m_a^2/E^2$.
Equation (\ref{E})
leads to the approximations
$
p_{a}
\simeq
E
+
\left( \rho - 1 \right)
m_{a}^2/2E
$,
$
v_{a}
\simeq
1
-
m_{a}^2/2E^2
$
and
$
\Omega_{a}
\simeq
2 \omega \sigma_x^2
$,
with
\begin{eqnarray}
&&
\sigma_x^2
\equiv
\sigma_{xP}^2 + \sigma_{xD}^2
\,,
\label{sigmax}
\\
&&
\omega
\equiv 1
+
\frac{\sigma_{xP}^2}{\sigma_x^2} \,
\frac{ \left( 1 - \vec{\ell} \cdot \vec{v}_{P} \right)^2 }
     { \lambda_P }
+
\frac{\sigma_{xD}^2}{\sigma_x^2} \,
\frac{ \left( 1 - \vec{\ell} \cdot \vec{v}_{D} \right)^2 }
     { \lambda_D }
\,.
\label{omega}
\end{eqnarray}
From Eq.(\ref{Sa})
one can see that in the relativistic approximation
$S_{a}(E_a)$
is minimum for
$ \vec{p}_P \simeq - \vec{p}_D \simeq E \vec{\ell} $
and
$ E_P \simeq - E_D \simeq E $.
These are the values of the kinematical parameters
for which the probability (\ref{exact})
of the process (\ref{01})
is not negligibly small
(without considering the geometrical suppression factor
$1/L^2$).
Hence,
energy-momentum conservation is approximately satisfied
in order to guarantee the observability
of the process (\ref{01})
and this approximate conservation
determines the value of the effective neutrino energy $E$.
From Eq.(\ref{Sa})
one can also see that
$\rho$
is a dimensionless quantity of order unity.
On the other hand,
$\omega$
could be rather large if
$ \lambda_{P(D)} $
is small.
For example,
if the production process is pion decay at rest with
$\alpha=\mu$,
$ \lambda_D \sim 1 $
and
$ \sigma_x \sim \sigma_{xP} $,
we have
$ \omega \sim 10 $.

In the relativistic approximation
one can factorize out of the sum over the mass eigenstate
indices $a$ and $b$
in Eq.(\ref{exact})
all the quantities that do not vanish
in the zeroth order of the expansion in powers of
$m_a^2/E^2$
(\emph{e.g.}
$\mathcal{A}_{a}$
and
$S_{a}\left(E_a\right)$).
Therefore,
in the relativistic approximation
for the flavor-changing probability
we have
\begin{equation}
P_{\alpha\beta}(\vec{L})
=
\sum_{a,b}
\mathcal{U}_{{\alpha}a}^{*} \,
\mathcal{U}_{{\beta}a} \,
\mathcal{U}_{{\alpha}b} \,
\mathcal{U}_{{\beta}b}^{*} \,
\exp\left[
- 2 \pi i
\frac{ L }{ L^{\mathrm{osc}}_{ab} }
-
\left( \frac{ L }{ L^{\mathrm{coh}}_{ab} } \right)^2
-
2 \pi^2 \omega \rho^2
\left( \frac{ \sigma_{x} }{ L^{\mathrm{osc}}_{ab} } \right)^2
\right]
\,,
\label{P}
\end{equation}
%
%
with
the oscillation lengths $ L^{\mathrm{osc}}_{ab} $
and the coherence lengths $ L^{\mathrm{coh}}_{ab} $,
for $a{\neq}b$,
given by
\begin{eqnarray}
&&
L^{\mathrm{osc}}_{ab}
\equiv
\frac{ 4 \pi E }{ \Delta{m}^2_{ab} }
\,,
\label{Losc}
\\
&&
L^{\mathrm{coh}}_{ab}
\equiv
2 \sqrt{2\omega}
\,
\frac{ 2 E^2 }{ \left| \Delta{m}^2_{ab} \right| } 
\, \sigma_{x}
\,,
\label{Lcoh}
\end{eqnarray}
where
$ \Delta{m}^2_{ab} \equiv m_a^2 - m_b^2 $.

The transition probability (\ref{P}) has the same form
as that given in Eq.(23) of \cite{GKLL93},
which was obtained in the framework of quantum field theory with
wave packets.
However,
the coefficients that
appear in the expression of the coherence length
and in front of the third term
in the exponential in Eq.(\ref{P})
are different from those of \cite{GKLL93}.
There are two reasons for these differences:
1) in \cite{GKLL93}
we assumed  the same spatial width
for all the wave packets
of the external particles involved in the process (\ref{01}),
whereas here these widths can be different;
2) in \cite{GKLL93}
we integrated over the momenta of the final particles,
whereas here we assume these momenta to be measured.

The first term in the exponential in Eq.(\ref{P})
is the usual oscillating phase
which gives rise to neutrino oscillations.
The second term causes a quadratical decrease of
$P_{\alpha\beta}(\vec{L})$
with the distance $L$
and determines how far the oscillations take place.
For $ L \gtrsim L^{\mathrm{coh}}_{ab} $
this term suppresses the interference of
the neutrino mass eigenstates $\nu_{a}$ and $\nu_{b}$,
leading to the disappearance of the oscillations
due to $L^{\mathrm{osc}}_{ab}$
for
$ L \gg L^{\mathrm{coh}}_{ab} $.
If $ L \gg L^{\mathrm{coh}}_{ab} $ for all $a{\neq}b $,
all the oscillating terms in the probability (\ref{P}) are suppressed,
leading to a constant transition probability
$
P_{\alpha\beta}
=
\sum_{a}
\left| \mathcal{U}_{{\alpha}a} \right|^2
\left| \mathcal{U}_{{\beta}a} \right|^2
$.
The third term in the exponential in Eq.(\ref{P}),
which is due to the time integration,
\emph{i.e.}
to the lack of time measurements,
implies that the interference terms are also washed out
if $\sigma_x$
is larger than the neutrino oscillation length.

Apart from the factor $ 2 \sqrt{2\omega}$,
the coherence length given in Eq.(\ref{Lcoh})
is similar to that obtained by physical intuition
in \cite{Nussinov,BP78,Kayser}
and in a quantum mechanical wave-packet treatment in \cite{GKL91}.
But there is a very important difference
between our result and the previous
formulas for the coherence length.
The coherence length is
usually defined as the distance at which the separation
between the wave packets of different massive neutrinos 
propagating with group velocities
$ v_{a} \simeq 1 - m_{a}^2 / E^2 $
is equal to $\sigma_{x}$,
\emph{i.e.}
\begin{equation}
L^{\mathrm{coh}}_{ab}
\equiv
\frac{ \sigma_{x} }{ \left| v_{a} - v_{b} \right| }
=
\frac { 2 E^2 }
      { \left| \Delta{m}^2_{ab} \right| }
\,
\sigma_{x}
\,,
\label{coh}
\end{equation}
where $\sigma_x$ is
the width of the propagating neutrino wave packet,
which
depends on the neutrino production mechanism.
However,
our calculations show that
the proper definition of the $\sigma_x$
that determines the coherence length
must also include information on
the neutrino detection mechanism.
We presented a rigorous definition of
$ \sigma_x $ in
Eq.(\ref{sigmax}).
Defining
$\sigma_p$
with the relation
$ \sigma_x \, \sigma_p = 1/2 $,
we have
\begin{equation}
\frac{1}{\sigma_p^2}
\equiv \frac{1}{\sigma_{pP}^2} + \frac{1}{\sigma_{pD}^2}
\,.
\label{sigmap}
\end{equation}
Hence,
it is clear that the
smaller of $\sigma_{pP}$ and $\sigma_{pD}$
dominates the value of $\sigma_p$.
If
$\sigma_{pD}$
is much smaller than
$\sigma_{pP}$,
then
$ \sigma_p \simeq \sigma_{pD} $
and
$ \sigma_x \simeq \sigma_{xD} $.
Therefore,
a precise measurement of the momentum
of all the particles involved
in the neutrino detection process\footnote{
It is necessary to measure accurately the
momentum of all the particles because
from Eq.(\ref{sigmaxP})
we have
$
\sigma_{pD}^2
=
\sigma_{pD_{I}}^2
+
\sigma_{pD_{F}}^2
+
\sigma_{p\beta}^2
$.
}
implies a very small
$ \sigma_p \simeq \sigma_{pD} $,
leading to a very large
$ \sigma_x \simeq \sigma_{xD} $
and a very large coherence length.
Thus,
our formulation with wave packets confirms
the observation in \cite{KNW96}
and provides a simple method to estimate
the $\sigma_x$
that determines the coherence length
through Eq.(\ref{Lcoh}):
$
\sigma_x
\simeq
\sqrt{ \sigma_{xP}^2 + \sigma_{xD}^2 }
$,
where
$\sigma_{xP}$ and $\sigma_{xD}$
are the estimated sizes of the production
and detection processes,
respectively.

The presence of the factor $ 2 \sqrt{2\omega} $
in the expression (\ref{Lcoh})
of the coherence length
can imply a non-negligible increase of
$L_{ab}^{\mathrm{coh}}$
with respect to the usual definition (\ref{coh}).
For example,
as we have already mentioned,
if the production process is pion decay at rest with
$\alpha=\mu$
and $ \lambda_D \sim 1 $,
we have
$ \omega \sim 10 $,
which gives
$ 2 \sqrt{2\omega} \sim 10 $.
Since $\omega$ is large if
$ \lambda_{P(D)} $
is small,
\emph{i.e.}
if all the particles involved
in the production (detection)
process have a very low velocity,
this increase of the coherence length
is due to the fact that the overlap of the wave packets
of slow particles last longer
and there is more time available for the production (detection)
process to emit (absorb) coherently the intermediate
superposition of massive neutrinos.
Indeed, one can estimate, for example,
that the time
$ \Delta{t}_P $
available for the production process is of the order of
$ \sigma_{xP}/|\vec{v}_P| $
and,
if $|\vec{v}_P|$ is small,
$ \lambda_D \sim 1 $
and
$ \sigma_x \sim \sigma_{xP} $,
we have
$
\omega
\sim
1 / |\vec{v}_P|^2
\sim
( \Delta{t}_P / \sigma_{x} )^2
$.
Hence, one can see that a small
$|\vec{v}_P|$
implies a large
$\Delta{t}_P$,
a large $\omega$
and a large coherence length.

On the other hand,
it must be noticed that
a large value of
$ \omega $
increases the contribution of
the third term in the exponential in the transition probability (\ref{P})
and
could enhance its suppression of the interference terms
if
$ \sigma_x \sim L^{\mathrm{osc}}_{ab} $.
This is due to the fact that when
the time available for the production (detection) process is longer
than
$L^{\mathrm{osc}}_{ab}$,
the interference of the neutrino mass eigenstates
$\nu_a$ and $\nu_b$ is washed out.
For example,
if
$
\omega
\sim
( \Delta{t}_P / \sigma_{x} )^2
$
as in the example above,
we have
$
\omega \, ( \sigma_x / L^{\mathrm{osc}}_{ab} )^2
\sim
( \Delta{t}_P / L^{\mathrm{osc}}_{ab} )^2
$
and the corresponding oscillating term in (\ref{P})
is suppressed if
$ \Delta{t}_P \gg L^{\mathrm{osc}}_{ab} $.

In summary,
the Grimus--Stockinger theorem considerably
simplifies the wave packet treatment.
The neutrino oscillations due to
$L^{\mathrm{osc}}_{ab}$
disappear
when the distance $L$ between neutrino production
and detection
is much larger than the coherence length
$L_{ab}^{\mathrm{coh}}$ given by Eq.(\ref{Lcoh})
with Eq.(\ref{sigmax}),
which is the most complete expression derived so far.
The authors of \cite{KNW96} pointed out that
one can observe neutrino oscillations
beyond the usually defined coherence distance
by a precise measurement of neutrino energy.
Our calculation confirms their claim
in the framework of a quantum field theoretical
calculation.
Another result of our calculation is that,
because of the lack of time measurements,
the neutrino oscillations due to
$L^{\mathrm{osc}}_{ab}$
can be observed only if
$\sigma_x$
is smaller than
$L^{\mathrm{osc}}_{ab}$.
Otherwise,
the interference of the mass eigenstates
$\nu_a$ and $\nu_b$
is washed out.

\acknowledgments
C.G. would like to thank W. Grimus for very useful discussions.
C.G. and U.W.L. would like to thank the
Korea Institute for Advanced Study
for their support and warm hospitality.
This work is supported in part by the Ministry of Education through
the Basic Science Research Institute, Contract No. BSRI-96-2418(UWL).


\begin{references}

\bibitem{Nussinov}
S. Nussinov,
Phys. Lett. B \textbf{63}, 201 (1976).

\bibitem{BP78}
S.M. Bilenky and B. Pontecorvo,
Phys. Rep. \textbf{41}, 225 (1978).

\bibitem{Kayser}
B. Kayser,
Phys. Rev. D \textbf{24}, 110 (1981).

\bibitem{GKL91}
C. Giunti, C.W. Kim and U.W. Lee,
Phys. Rev. D \textbf{44}, 3635 (1991).

\bibitem{GKLL93}
C. Giunti, C.W. Kim, J.A. Lee and U.W. Lee,
Phys. Rev. D \textbf{48}, 4310 (1993).

\bibitem{Rich}
J. Rich,
Phys. Rev. D \textbf{48}, 4318 (1993).

\bibitem{KNW96}
K. Kiers, S. Nussinov and N. Weiss,
Phys. Rev. D \textbf{53}, 537 (1996).

\bibitem{GS96}
W. Grimus and P. Stockinger,
Phys. Rev. D \textbf{54}, 3414 (1996).

\bibitem{Campagne}
J.E. Campagne,
Phys. Lett. B \textbf{400}, 135 (1997).

\end{references}
\end{document}